\documentclass[fdp,a4paper,fleqn%
]{w-art}
\usepackage{times,cite,w-thm}
\theoremstyle{plain}

\theoremstyle{definition}

\usepackage[]{graphicx}
\begin{document}
\Volume{55}
\Month{01}
\Year{2007}
\pagespan{1}{}



\title[The ``landscape" of  Pati--Salam heterotic superstring vacua]{The ``landscape" of Pati--Salam heterotic superstring vacua}


\author[J. Rizos]{J. Rizos\inst{1,}%
  \footnote{
Presented at School and Workshops on Cosmology and Strings: Theory-Cosmology-Phenomenology (CORFU2009), September 6-13, Corfu , Greece.}}
\address[\inst{1}]{Theory Division, Physics Department, University of Ioannina, 45110 Ioannina, Greece}
\begin{abstract}
 The main aspects of a recently developed method for classification of heterotic superstring vacua in
 the Free Fermionic Formulation are presented.
 Using these techniques we classify a big number of approximately $10^{15}$ heterotic string vacua with Pati--Salam,
 $SU(4)\times{SU(2)}_L\times{SU(2)}_R\,$, gauge symmetry with respect to their main phenomenological
 features as the number of families, Pati--Salam breaking Higgs,
  Standard Model Higgs doublets, additional triplets and exotic charge states.
  We identify an interesting subclass of these vacua, approximately one to one million,
   whose massless spectrum is completely free of fractionally charge states.
\end{abstract}
\maketitle                   

\section{Introduction}
String theory provides a framework for unification of gauge interactions with gravity.
There exist only five string theories in 10 dimensions, the $E_8\times{E_8}$ and $SO(32)$ heterotic, the
type I, type IIA and type IIB strings. However, compactification leads to a plethora of string vacua in four dimensions.
Recently, a lot of effort has been devoted in studying the properties of these vacua, also referred as ``string landscape"
\cite{land}, including  intersecting D-brane orientifold models \cite{tyiim} and heterotic models \cite{hetclass,ffm}.

In references \cite{ttc,ffm} a new approach has been developed that allows to study
 a big class of heterotic vacua,  with $SO(10)$ embedding, in the context of Free Fermionic Formulation \cite{fff}.
A model in this class is defined by a set of 12 basis vectors
\begin{align}
v_1=1&=\{\psi^\mu,\
\chi^{1,\dots,6},y^{1,\dots,6},\omega^{1,\dots,6}|\bar{y}^{1,\dots,6},
\bar{\omega}^{1,\dots,6},\bar{\eta}^{1,2,3},
\bar{\psi}^{1,\dots,5},\bar{\phi}^{1,\dots,8}\}\nonumber\\
v_2=S&=\{\psi^\mu,\chi^{1,\dots,6}\}\nonumber\\
v_{2+i}=e_i&=\{y^{i},\omega^{i}|\bar{y}^i,\bar{\omega}^i\}, \ i=1,\dots,6\nonumber\\
v_{9}=b_1&=\{\chi^{34},\chi^{56},y^{34},y^{56}|\bar{y}^{34},\bar{y}^{56},
\bar{\eta}^1,\bar{\psi}^{1,\dots,5}\}\label{bas}\\
v_{10}=b_2&=\{\chi^{12},\chi^{56},y^{12},y^{56}|\bar{y}^{12},\bar{y}^{56},
\bar{\eta}^2,\bar{\psi}^{1,\dots,5}\}\nonumber\\
v_{11}=z_1&=\{\bar{\phi}^{1,\dots,4}\}\nonumber\\
v_{12}=z_2&=\{\bar{\phi}^{5,\dots,8}\}\nonumber
\end{align}
and a set of  66 phases $c[v_i,v_j]=\pm1, j<i=1,\dots,12$, that correspond to generalized GSO projections (GGSO). The
basis vectors $b_1,b_2$ correspond to $Z_2\times Z_2$ orbifold twists, while $e_i,i=1,\dots,6$ correspond to shifts.
Keeping the basis vectors fixed and varying the GGSO phases we can generate  $2^{12(12-1)/2}=2^{66}$ models
with $SO(10)\times{U(1)}^3\times SO(8)^2$ gauge symmetry. Following \cite{ffm} we can derive expressions for the
main phenomenological aspects of the model in terms of the GGSO phases. The number of twisted spinorial representations, per orbifold
plane $I=1,2,3$,
is given by
\begin{align}
\#(S^{(I)})=\left\{
\begin{array}{ll}
2^{4-{\rm rank}\left(\Delta^{(I)}\right)}&{\rm rank}\left(\Delta^{(I)}\right)={\rm rank}\left[\Delta^{(I)},Y_{16}^{(I)}\right]
\nonumber\\
0&{\rm rank}\left(\Delta^{(I)}\right)<{\rm rank}\left[\Delta^{(I)},Y_{16}^{(I)}\right]\nonumber
\end{array}
\right.
\end{align}
while the number of vectorial representations is given by
\begin{align}
\#(V^{(I)})=\left\{
\begin{array}{ll}
2^{4-{\rm rank}\left(\Delta^{(I)}\right)}&{\rm rank}\left(\Delta^{(I)}\right)={\rm rank}\left[\Delta^{(I)},Y_{10}^{(I)}\right]
\nonumber\\
0&{\rm rank}\left(\Delta^{(I)}\right)<{\rm rank}\left[\Delta^{(I)},Y_{10}^{(I)}\right]\nonumber
\end{array}
\right.
\end{align}
where
$\Delta^{(I)}$ are 4x4 and $Y^{(I)}$  are 4x1 GGSO coefficient matrices and $\left[\Delta^{(I)},Y_{10}^{(I)}\right],
\left[\Delta^{(I)},Y_{16}^{(I)}\right]$ represent the enchased 4x5 matrices. Similar expressions can be derived for the chirality
of spinorial representations. Using these results a full classification of  $SO(10)$ vacua, numbering approximately
 $10^{20}$ models, has been presented
in \cite{ffm}, revealing some interesting properties of the spectra as the spinor-vector duality.

\section{The Pati--Salam superstring ``landscape"}
In this work we report some recent progress in extending the analysis of \cite{ffm} by adding an extra basis vector to \eqref{bas} that
breaks the $SO(10)$ symmetry. The simplest choice is the vector
\begin{align}
v_{13}=\alpha=\{\bar{\psi}^{45},\bar{\phi}^{1,2}\}
\end{align}
that introduces 12 new  GSO projection phases $c[\alpha,v_j],j=1,\dots,12$. The full gauge group (away from special points) is broken down to
$SU(4)\times{SU(2)}_L\times{SU(2)}_R\times{U(1)}^3\times{SU(2)}^4\times{SO(8)}$ which includes the Pati--Salam as the
symmetry of the ``observable" sector \cite{ps}.

A heterotic string realization of a supersymmetric version of the  Pati--Salam model has been studied in \cite{pss}. The SM particles
are accommodated in PS multiplets

$SU(4)\times{SU(2)}_L\times{SU(2)}_R\supset SU(3)\times{SU(2)}\times{U(1)}$
\begin{align}
    {F}_L({\bf4},{\bf2},{\bf1})     &\rightarrow
     Q({\bf3},{\bf2},-\frac 16) + \ell{({\bf1},{\bf2},\frac 12)}
     \nonumber\\
\bar{F}_R({\bf\bar 4},{\bf1},{\bf2})&\rightarrow  u^c({\bf\bar 3},
{\bf1},\frac 23)+d^c({\bf\bar 3},{\bf1},-\frac 13)+
                            e^c({\bf1},
                           {\bf1},-1)+\nu^c({\bf1},{\bf1},0)
                           \nonumber\\
D({\bf6},{\bf1},1)  &\rightarrow   D_3({\bf3},{\bf1},\frac 13) +
\bar{D}_3({\bf\bar 3},{\bf1},-\frac 13)
                             \nonumber\\
h({\bf1},{\bf2},{\bf2})
 &\rightarrow   h^d({\bf1},{\bf2},\frac 12) + h^u({\bf1},{\bf2},-\frac 12)
 \nonumber
\end{align}
where $D_3,\bar{D}_3$ are additional quark triplets. Using the standard hypercharge embedding, that guarantees $sin^2\theta_w=\frac{3}{8}$
at the string scale, $Y=\frac{1}{\sqrt{16}}\,T_{15}+\frac{1}{2}\,I_{3L}+\frac{1}{2}\,I_{3R}$,
states transforming as $({\bf1},{\bf1},{\bf2})$, $({\bf1},{\bf2},{\bf1})$, $({\bf4},{\bf1},{\bf1})$ accommodate fractional charge exotics.
The appearance of these states is a generic feature of $k=1$ string compactifications as shown in \cite{fra}.

The PS symmetry can be broken to the Standard Model by vevs to the neutral components ($\left<\nu^c_H\right>,\left<\nu_H\right>$) of
a pair of Higg multiplets ($H,\bar{H}$)
\begin{align}
\bar{H}({\bf\bar 4},{\bf1},{\bf2})&\rightarrow   u^c_H({\bf\bar 3},
{\bf1},\frac 23)+d^c_H({\bf\bar 3},{\bf1},-\frac 13)+
                            \nu^c_H({\bf1},{\bf1},0)+
                             e^c_H({\bf1},{\bf1},-1)
                             \nonumber\\
{H}({\bf4},{\bf1},{\bf2})&\rightarrow  u_H({\bf3},{\bf1},-\frac
23)+d_H({\bf3},{\bf1},\frac 13)+
              \nu_H({\bf1},{\bf1},0)+ e_H({\bf1},{\bf1},1)
\end{align}
The triplet remnants of the Higgs mechanism get heavy masses combined with the triplets in $D({\bf6},{\bf1},1)$ through the
superpotential terms
\begin{align}
H^2\,D+\bar{H}^2\,D\to d_H\,\bar{D}_3\left<\nu_H\right>+{d}^c_H\,{D}_3\,\left<\nu^c_H\right>
\end{align}
Fermion masses are generated by the superpotential term
\begin{align}
{F}_L({\bf4},{\bf2},{\bf1})\bar{F}_R({\bf\bar 4},{\bf1},{\bf2})\,h({\bf1},{\bf2},{\bf2})
\end{align}
however neutrinos stay naturally light as they mix with additional heavy singlets \cite{pss}.

\begin{figure}[!h]
\includegraphics[width=8.7cm]{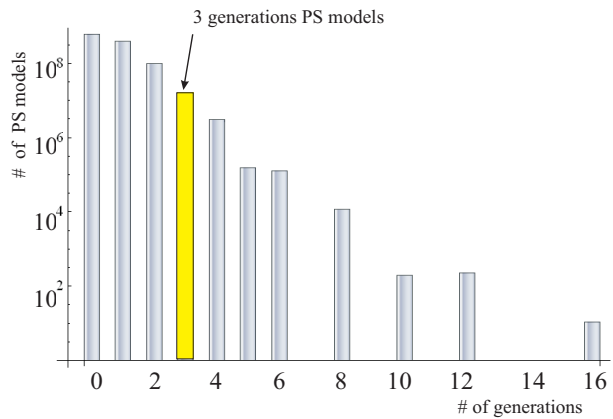}%
\caption{Number PS models versus number of generations in a random sample  of $5\times10^9$ models.}
\label{fig:gen}
\end{figure}

\begin{figure}[!h]
\includegraphics[width=9cm]{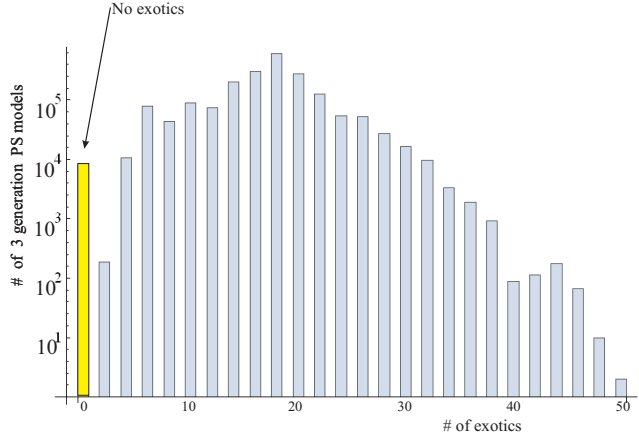}%
\caption{Number 3 generation PS models versus number number of exotic state multiplets in a random sample  of $5\times10^9$ models.}
\label{fig:gent}
\end{figure}

Following the method developed in \cite{ffm}, analytic formulae are derived for the main features of a PS model massless spectrum in terms
of GGSO coefficients. These include
the number of ${F}_L, \bar{F}_L, {F}_R, D, \bar{D}, H$, $\bar{H}$, $h$ as well as the numbers of exotic multiplets. After fixing some coefficients to
guarantee $N=1$ supersymmetry, and removing some redundancy,  we are left with 51 independent phases giving rise to a class of $2^{51}\sim 10^{15}$ vacua.
 Using a
computer program and random sampling over $5\times 10^9$ models we obtain a statistical profile of this class of models. In Figure \ref{fig:gen}
we plot the number of PS models versus the number of fermion generations. We remark that 3-generation vacua  correspond
approximately to 0.3\% of the sample. Furthermore, in Figure \ref{fig:gent}, we  plot the number of models versus the number
of exotic multiplets in the spectrum. We observe that an important number of models in the sample, approximately one PS model per half milion,
is free of massless exotics. This phenomenologically interesting result is not in contradiction with \cite{fra},
since exotics are still present in the massive
heterotic string spectrum. A detail  analysis
of the ``landscape" of PS heterotic vacua is presented in \cite{exo}.
An outline of the main results is shown in Table \ref{tab:one}, where we examine a series of phenomenological constraints
and calculate the statistics of the models
satisfying them. Every line in the table corresponds to an additional constraint. We remark that the number of vacua satisfying all requirements,
is approximately one to one million. This reveals a rich subclass of phenomenologically interesting vacua that deserve further study.


\begin{table}[<float>]
\begin{tabular}{clcc}
\hline
\hline
&Constraint&probability& estimated \# of models\\
\hline
\hline
&No gauge group enhancements & $2\times10^{-1}$&$2\times10^{14}$\\
\hline
+&3 generation models  & $3\times10^{-3}$&$7\times10^{12}$\\
\hline
+&PS breaking Higgs & $4\times10^{-4}$&$9\times10^{11}$\\
\hline
+&SM breaking Higgs doublets& $3\times10^{-4}$&$7\times10^{11}$\\
\hline
+&No exotics &$1\times10^{-6}$&$2\times10^9$\\
\hline
\end{tabular}
\caption{PS models statistics with respect to phenomenological constraints imposed on massless spectrum.}
\label{tab:one}
\end{table}
\begin{acknowledgement}
This work is supported in part by the EU under contract MRTN–CT–2006–035863–1.
The author would like to thank University of Liverpool and CERN Theory Division for the hospitality.
\end{acknowledgement}


\end{document}